\documentclass{sigchi}

\CopyrightYear{2020}
\setcopyright{acmlicensed}
\doi{https://doi.org/10.1145/3313831.XXXXXXX}
\isbn{978-1-4503-6708-0/20/04}
\conferenceinfo{CHI'20,}{April  25--30, 2020, Honolulu, HI, USA}
\acmPrice{\$15.00}

\usepackage{balance}       
\usepackage{graphics}      
\usepackage[T1]{fontenc}   
\usepackage{txfonts}
\usepackage{mathptmx}
\usepackage[pdflang={en-US},pdftex]{hyperref}
\usepackage{color}
\usepackage{booktabs}
\usepackage{textcomp}
\usepackage{cuted}
\usepackage{capt-of}

\usepackage{microtype}        
\usepackage{ccicons}          

\usepackage{todonotes}

\def\plaintitle{SIGCHI Conference Proceedings Format}

\def\emptyauthor{}
\def\plainkeywords{Authors' choice; of terms; separated; by
  semicolons; include commas, within terms only; this section is required.}

\makeatletter
\def\url@leostyle{%
  \@ifundefined{selectfont}{
    \def\UrlFont{\sf}
  }{
    \def\UrlFont{\small\bf\ttfamily}
  }}
\makeatother
\def\pprw{8.5in}
\def\pprh{11in}

\definecolor{linkColor}{RGB}{6,125,233}
\hypersetup{%
  pdftitle={\plaintitle},
  pdfauthor={\emptyauthor},
  pdfkeywords={\plainkeywords},
  pdfdisplaydoctitle=true, 
  bookmarksnumbered,
  pdfstartview={FitH},
  colorlinks,
  citecolor=black,
  filecolor=black,
  linkcolor=black,
  urlcolor=linkColor,
  breaklinks=true,
  hypertexnames=false
}

\begin{document}

\title{CoAug-MR: An MR-based Interactive Office Workstation Design System via Augmented Multi-Person Collaboration}

\numberofauthors{2}
\author{%
  \alignauthor{Lin Wang\\
    \affaddr{Visual Intelligence Lab.}\\
    \affaddr{KAIST}\\
    \email{wanglin@kaist.ac.kr}}\\
  \alignauthor{Kuk-Jin Yoon\\
     \affaddr{Visual Intelligence Lab.}\\
     \affaddr{KAIST}\\
    \email{kjyoon@kaist.ac.kr}}\\
}
\maketitle
\begin{strip}
    \centering
    \includegraphics[width=\textwidth,height=4cm]{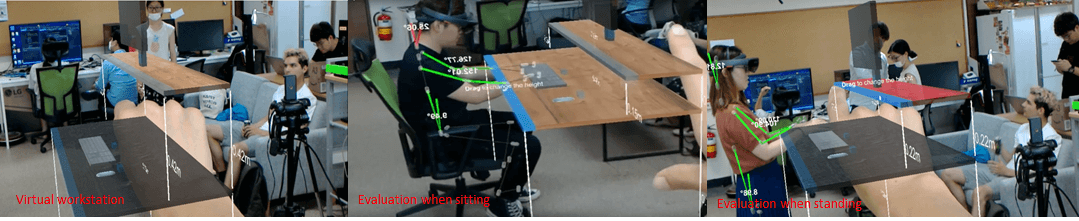}
    \captionof{figure}{ This paper proposes CoAug-MR, a MR-based interactive design method via augmented multi-person collaboration. We emphasize the importance of collaboration and initiate an augmented design paradigm, which enables user, designer, and even ergonomist to gather up to achieve reliable and real-time design and prototyping based on user's pose either locally or remotely.
}
    \label{fig:impressive} 
\end{strip}
\begin{abstract}
 Digital human models (DHMs) are useful for designing products considering the general population. Besides, the Do-It-Yourself (DIY) approach somehow enables fast prototyping by casual users themselves via virtual mannequins. However, DHMs suffer from insufficient body data, and DIY lacks reliability and quality of design due to a lack of professional guidance and less accessibility of 3D modeling tools. This paper addresses the above-mentioned problems and proposes CoAug-MR, an MR-based interactive design method via augmented multi-person collaboration. We emphasize the importance of collaboration and initiate an augmented design paradigm, which enables user, designer, and ergonomist to gather up to achieve reliable and real-time design and prototyping based on user's pose either locally or remotely. Our system consists of an information-sharing module, a multi-person interaction module, and a spatial engagement module. Usability study shows that our augmented collaborative design system is effective for reliable and fast product design and evaluation.
\end{abstract}




\section{Introduction}

In recent years, digital modeling and prototyping have been burgeoning in the field of industrial design and human-computer interaction (HCI). Computer Aided Design (CAD) tools are typical platforms that focus on the virtual modeling of 2D and 3D products. These tools are more utilized by professionals who have mastered the knowledge of creating 3D geometry through the CAD interface. However, some tools are relatively difficult to learn, and special training is needed for familiarization of the interface. Hence, research has been concentrated on creating novel interfaces using gestures, sketching \cite{saul2011sketchchair}, virtual reality \cite{peng2007virtual}, augmented reality \cite{morisawa2003development}, and other existing objects \cite{follmer2010copycad} as references. On the other hand, digital human models(DHMs) are the other type of tools that enable ergonomic evaluation for the quality of products. DHMs are utilized to assess the safety and usability of products and environment in the virtual space \cite{park2008design}. However, it suffers from the difficulties of accurate prediction of human behaviors due to the lack of human anthropometric data. Most DHMs tend to represent variations of body shapes for the general population, which can be a challenging problem when it comes to customizing products. Now, the question is how to make the product design connected to those people who are not only in the general range but also beyond the general range? 
One challenge is how to connect a user's anthropometrics with pose recommendations and design dimensions, which may take large amount of time to get information from an individual in the usual way. Another challenge is the ergonomic evaluation for the individuals out of the general range. Since most of the ergonomic evaluations using DHMs have highly relied on the general population, it makes the evaluations inapplicable to these minorities. 

With these challenges ahead,  research has been focused on real-time ergonomic evaluations. Recent research endeavors to use sensor tracking devices \cite{schumann1998applying} and wire-free sensor-based device \cite{pastura2013joint} to provide an ergonomic guideline for the individuals. Since these devices support real-time interaction with objects, and they provide a way to get body information through sensors and trackers. Sensors can be integrated into modern computer graphics technology such as Virtual Reality (VR). It also helps to offer new possibilities when we perform product design \cite{ottosson2002virtual}, design evaluation \cite{ye2007applications}, ergonomic evaluation \cite{grecco2007virtual,lee2018interactive}, simulation \cite{ikonomov2003using} and manufacturing \cite{berg2017industry}. It turned out that some VR technologies aiming to design, simulation and manufacturing seem to be mature, but, in order to be more scientific, human-centered VR system of product evaluation still needs a long way to go.  The reason why ergonomic evaluation for human-centered systems is difficult is that humans vary from one to another and a certain range of body parts cannot represent others.  In a VR environment, human bodies are replaced by virtual mannequins, which actually impede the intuitive interaction with the 3D world in the evaluation of human-centered systems. Some research utilizes the body scan and 3D reconstruction to make an avatar for individuals \cite{grecco2007virtual}, however, it seems to be very cumbersome and time-consuming when we have to scan and reconstruct each by each. On the other hand, some studies try to enable casual users themselves to design and evaluate body postures by themselves \cite{grecco2007virtual, lee2016posing}, in which the method is called Do-It-Yourself (DIY). Although the framework could be good in some way, since they know better what they want, and it could be more convenient for their personal preference, it cannot ensure reliability and quality.

Regarding these challenges and drawbacks, it makes us more conscious about how to make design possible for all body types through a better way. What if we do not need to try to scan the body and make an mannequin to make interaction with virtual products? What can we do to overcome the uncertainty and unreliability of DIY?  What can we do to make a more instant ergonomic evaluation and design feedback for all types of users? What if prototyping manifests itself in a way that designer, user, and even ergonomist could virtually gather together to accomplish fast prototyping? With these questions, in  this paper, we aim to accomplish reliable and real-time design and evaluation in physical space with mixed Reality (MR).   
Unlike VR, which creates an entirely new and immersive 3D environment, MR adds more information to the user's current environment. Meanwhile, in order to meet the prejudice of casual users and overcome the unreliability of the design through DIY, it would be better if we include designer and ergonomist to the design process. We emphasize the importance of collaboration, and initiate an augmented design paradigm, which enables user, designer and even ergonomist to gather up to achieve reliable and real-time design and prototyping based on user's pose either locally or remotely.  
The contributions of this paper are as follows: (1) A novel interactive design strategy for reliable and instant prototyping by connecting user, designer and ergonomists together. (2) CoAug-MR, a novel MR-based interactive design system that enables  multi-person collaboration locally or remotely. (3) An initial usability study for the design system that shows the effectiveness of the proposed method.
 \begin{table*}[t!]                  
 \begin{center}
  \begin{tabular*}{\textwidth}{p{2.2cm}|p{2.2cm}|p{2.2cm}|p{2.2cm}|p{2.2cm}|p{2.2cm}|p{2.2cm}}
    \hline
    \centering
    &   Weichel \cite{weichel2014mixfab} & Lee     \cite{lee2018interactive}& Taylor\cite{taylor2013posture}& Pontonnier   \cite{pontonnier2014designing}&  Saul \cite{saul2011sketchchair}  &Ours\\ \hline
    \centering
     Research focus &   DIY frabraction &  DIY ergonomic guidelines&Pose training &Ergonomic design&Chair sketch design system&    Augmented collaborative design       \\ \hline  \centering   
     Platform & AR &VR&VR&VR&VR&MR \\ \hline \centering
     H/W & Kinect/HMD&  Kinect/HMD& Kinect&Mocap, EMG& Webcam markers &Hololens, motion tracker,Kinect
\\ \hline \centering
     S/W & N/A &Unity 3D &N/A& N/A &NyARToolkit &Unity 3D
  \\ \hline \centering
    Augmented collaboration &No &No &No&No&No&Yes
  \\ \hline
    Real-time  ergonomic evaluation  &N/A&Yes&Yes&N/A&No&Yes
   \\ \hline \centering
    Active pose measuring &No&No &Yes &No&No &Yes \\ \hline \centering
    Interactive simulation in design &Yes  &Yes &No&No&Yes &Yes \\ \hline \centering
    Evaluation of system &Yes&Yes&No&No&Yes &Yes
 \\ \hline
    \end{tabular*}
    \end{center}
    \caption{A systematic comparison with the state-of-art works. Different from other works, we propose a novel MR-based design system based on augmented multi-person collaboration. Our method enables active pose measuring and interactive simulation in product design. Besides, through augmented collaboration among casual users and professionals, our method also facilitates real-time ergonomic evaluation, instant feedback and instant prototyping without the limitation of location. (N/A indicates not available.)  }
    \label{tab:msg1}                
\end{table*}
\section{Related works}
Designing interfaces or systems for modeling 3D objects that are accessible to casual users is of great importance for customization since they enable non-professionals to show their preferences, needs and expectations, meanwhile, engage themselves in design by themselves or with specialists. As mentioned in \cite{grecco2007virtual}, it is very crucial to provide people with the capacity to create what they want and make the design fit their geometries. They focus on building the experience prototyping technique initiated in \cite{wu2013usability}. However, how to know whether the pose is good or not should highly rely on ergonomic evaluation \cite{grecco2007virtual,pastura2013joint,haggag2013real}.   
\subsection{Understanding user and pose measuring} 
The intention of understanding users is that, in a user-centered approach, they need to model 3D products through design interfaces. Lee et al. \cite{lee2018interactive} conduct formative and evaluative studies that target on the casual users in the iterative process of personalizing items related to their use. In this way, casual users have the chance to determine dimensions using their body posing and acting. In the virtual environment, they can experience their design idea and evaluate their design through the feedback. However, the prototype captures the user's body pose and supports a referral of a wide range of body parts, which can only be working for some simple boxed based design. For more complex designs, the location and hand gestures and poses must be captured to support reference plane. 

On the other hand, understanding the user's pose is also a challenging task. Some studies have focused on using wire-embedded sensors to measure poses \cite{mattmann2007recognizing, wang2016zishi}, others use wireless sensors\cite{taylor2013posture} and augmented mirror\cite{anderson2013youmove} to measure poses.  Saul et al. in \cite{saul2011sketchchair} develop a design system for end users. The system allows end users to draw different styles of chairs and then a virtual human representing the user is placed on the chair to verify the stability of the chair. However, the DHMs cannot represent the real human body and the anthropometric information is limited due to a shortage of data. Therefore, in this research, we claim that using the real human body for evaluation should be the better and more reliable approach for end-users. In summary, these studies strive to help users to understand and retain comfortable poses in their work, however, they fail to take the work environment into account, which could be problematic when their poses are disjointed with their work circumstances. 
\subsection{Ergonomic evaluation in design }
Ergonomic evaluation is extremely crucial part of the design. Woldstad et al. \cite{wegner2007digital} use DHMs to design both equipment and work environments to meet human operations. Through the use of DHMs, designers and ergonomists can position and manipulate operator with different anthropometry within the simulated work environment.  Mochimaru et al. \cite{mochimaru2017digital} try to target people with special and embed digital human models to support design workshop. However, DHMs suffer from insufficient anthropometric and strength data, and the accuracy of prediction still needs to be improved to reach the goal of accurate prediction. In recent years, some real-time ergonomic assessment methods have provided a great platform for the design process. In \cite{haggag2013real}, Haggag et al. use Kinect sensor as a platform to support real-time rapid upper limb assessment (RULA) for assembly operations in industrial environments. They use voxel-based angle estimation to calculate angles of each joint of the upper body part. However, the joint occlusion makes the estimation unstable, and hand tracking is not supported in the current sensor, which impedes the arm assessment. In \cite{pastura2013joint}, Pastura et al. developed a computational system for joint angles calculation of human body. They use tracking markers positioned on specific anatomical points of the body to design define body segments. Through the AR environment, they build a low-cost infrastructure, which is comprised of webcam and paper makers to compute joint angles. Lee et al. \cite{lee2018interactive} then visualize the ergonomic guideline for furniture design.They explore ergonomic guidelines in personal fabrication and identify a dependency map between the user's anthropometrics, pose recommendations and design dimensions. 

In this paper, we explore to provide intuitive ergonomic evaluation and instant design prototyping based on user's on-site performance of tasks. One problem about general ergonomic guidelines for office workstation design is that they are most fit to the large percentage of an intended user population, and are confined with a range from 5 percentiles to 95 percentiles. On the other hand, as mentioned in \cite{lee2016posing}, the guidelines for ergonomists are based on the observation of body poses and angles to evaluate design quality, however, these guidelines are not directly aimed for individuals and for providing reliable design recommendations. As they have divided the guidelines into body-centered guidelines, object-centered guidelines and space-centered guidelines. Body clearance indicates the distance between body and desk, body reachability indicates the limitations of our hands and arms. They also modified the guideline from the general population to a person. The elbow angle for sitting posture is explained between 70 to 100 degrees. In this paper, we consider using this person-oriented ergonomic guideline for design evaluation and decision-making based on \cite{lee2016posing}.
\begin{figure*}[t!]
     \centering
     \includegraphics[width=\textwidth, height=6cm]{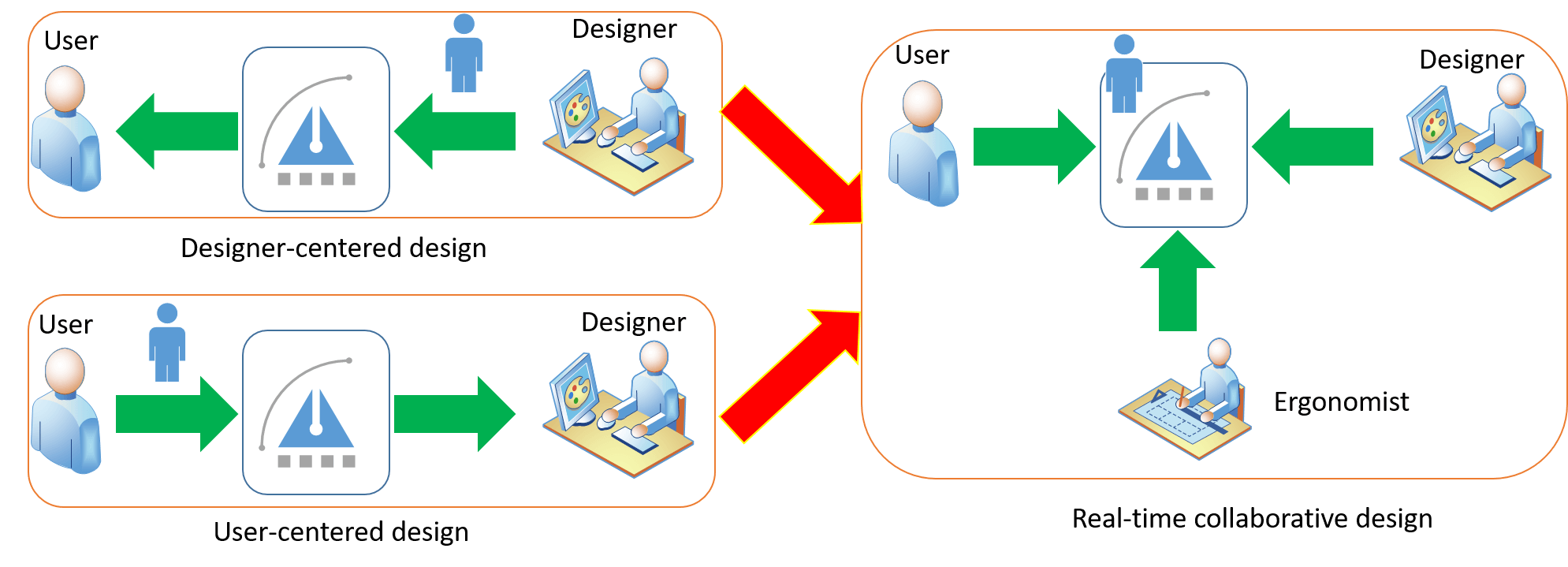}
     \caption{The proposed design strategy. Instead of focusing on designer-centered design or user-centered design, we initialize collaborative design among both casual users and specialists trough augmented collaboation, which enables fast and reliable prototyping and evaluation. }
     \label{fig:codesign}
 \end{figure*}
\subsection{Collaboration design system and VR/AR}
In \cite{yee2009augmented}, Yee et al. propose a method to do in-situ 3D sketching in the augmented environment. Gjosater et al. in \cite{saul2011sketchchair} present research called FurnitureAR, which applies augmented reality to collaborative furniture design. They highlight the idea of collaboration among designers who can first sketch design ideas in the CAD program and load model to FurnitAR for collaboration. In the collaborative environment, the CAD model can be freely modified by designers and, meanwhile, the usability factors can also be determined. However, the ergonomic evaluation of the design model is not considered. In this paper, we argue that human-centered design with a group of professionals and non-professionals is much more reliable and efficient than one casual user or designers themselves. The first collaborative design framework with ergonomists was initiated by Pontonnier et al. in \cite{pontonnier2014designing}. They propose to use a VR-based immersive virtual environment to represent workstation. They suggest to compute and visualize bio-mechanical factors that could be represented as virtual manikin or simple curves to ergonomist. The basic idea of the framework is to user sensor-bridging and sensor-sharing information to enhance the user's cognition and comprehension of the task. However, this idea was just considered as a potential application of the VR-supported prototyping of the workstation design. Their proposal can provide us the indication for collaborative design with professionals and end users. However, in  virtual environment, human body can only be represented by  virtual mannequins, which can not represent real human body. Thus, it is difficult to conduct accurate ergonomic evaluation in virtual environment.  
Noh et al.\cite{noh2015hmd} propose an MR system for avatar-mediated remote collaboration. They present an application that enables a local user to interact and collaborate with another user from remote space using natural hand motion. The system summons the remote user to the local space, which is displayed as a virtual avatar in the real world view seen by the user.  We claim that DIY design is not reliable than collaboration. We emphasize the importance of collaboration, and initiate an augmented design paradigm, which enables user, designer and even ergonomist to gather up to achieve reliable and real-time design and prototyping based on user's pose either locally or remotely. 
\begin{figure*}[t!]
    \centering
    \includegraphics[width=0.95\textwidth, height=6cm]{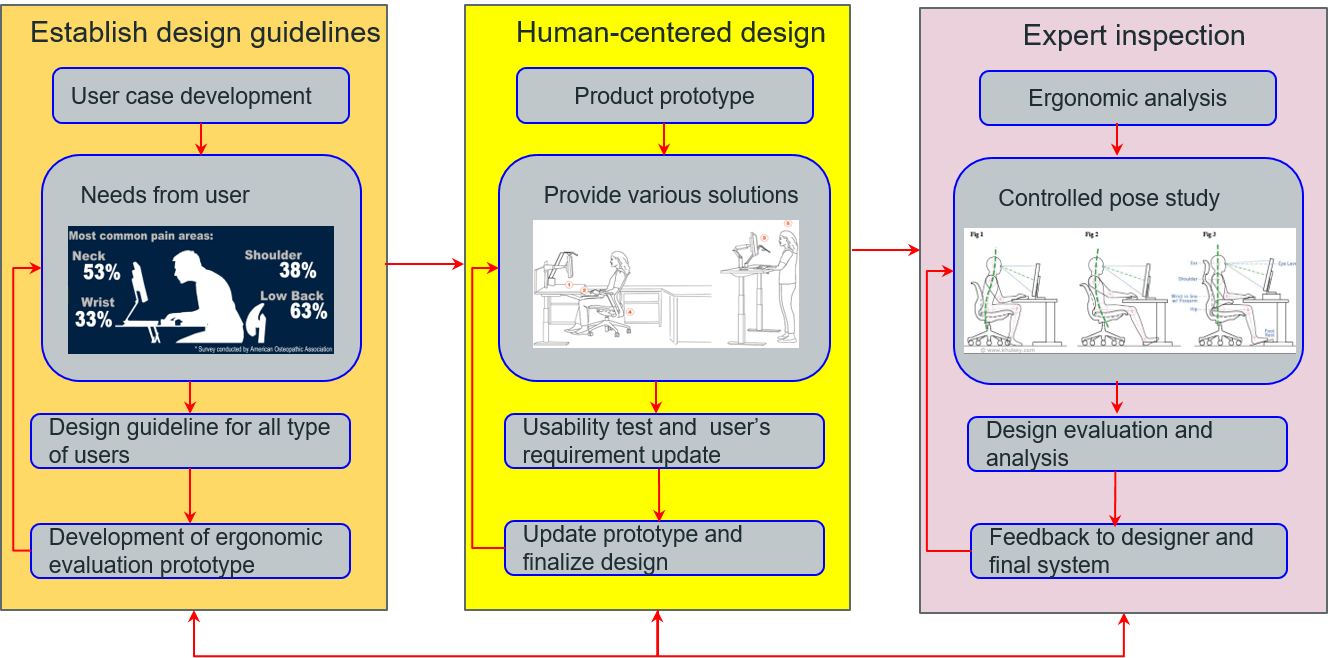}
    \caption{An illustration of our augmented collaboration design process. Our method incorporates the design guidelines based on the need from user, human-centered design paradigm (both user and designer) and expert participation and inspection (designer and ergonomist). With this strategy, the design system enables the local or remote collaboration for the design and evaluation among the user, designer and ergonomist.  }
    \label{fig:design_paradigm}
\end{figure*}

\begin{figure}[t!]
    \centering
    \includegraphics[width=\columnwidth, height=3cm]{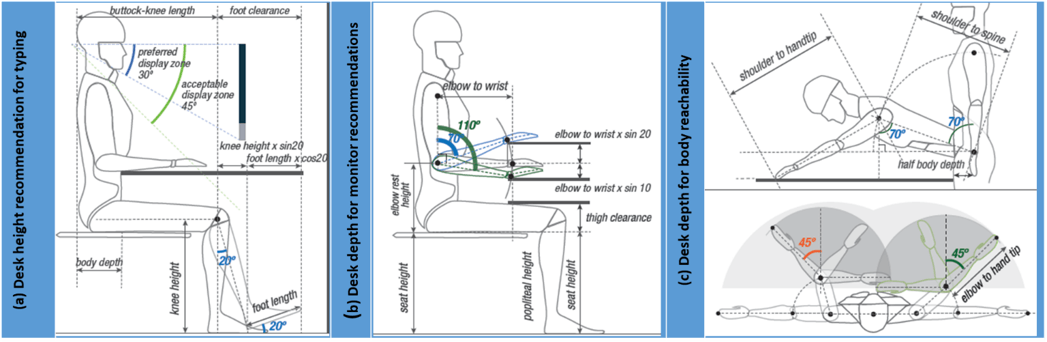}
    \caption{Person-centered ergonomic design guidelines. We adopt the ergonomic guidelines from [7].  }
    \label{fig:guideline}
\end{figure}
\section{Proposed Method}
In this section, we describe the methodology for our design system via augmented collaboration. \emph{For convenience, we focus on office workstation design as a typical example for our method.} There are some basic challenges to be solved for our system. \emph{First of all}, we try to overcome the drawbacks of capturing the human body as a virtual human model as done in \cite{lee2018interactive, pontonnier2014designing}. Since it takes too much time and energy to scan the body and make an avatar, we propose a new way to represent human body in the interactive environment. Our approach is to avoid avataring user but instead to directly utilize real body.  \emph{Second}, as discussed in previous section, DIY is limited and less reliable, thus we instead address the concept of collaborative design by which we get rid of the demerits of pure user-centered design and pure designer-centered design paradigms. As shown in Fig.~\ref{fig:codesign}, our method is to elaborate a platform where casual users and specialists can all perform their design and evaluation. Our goal is to enable designer and ergonomist from either local space or remote space to have collaboration with users to accomplish an efficient and reliable design. \emph{Third}, we aim to design more intuitive and situated experience according to the design guidelines. We aim to create more immersive design environment by using hand gestures, poses, and locations in MR environment. \emph{Last but not least}, we strive to create a platform that is flexible for the designer and ergonomist to participate in the design and ergonomic evaluation no matter where they are ( either locally or remotely). 
\subsection{Augmented collaborative design paradigm}
\subsubsection{Ergonomic guidelines for personal fit}
 As shown in the left portion of Fig.~\ref{fig:design_paradigm}, the designer needs to take care of the design of office workstation that should be able to prevent any repetitive strain injury, eye strain, fatigue, and discomfort. 
 One problem about general ergonomic guidelines for office workstation design is that they are most fit to the large percentage of an intended user population, and are confined with a range from 5 percentiles to 95 percentiles. On the other hand, as mentioned in \cite{lee2016posing}, the guidelines are based on the observation of body poses and angles to evaluate design quality, however, these guidelines are not directly aimed for individuals, thus making it hard for providing reliable design recommendations. As \cite{lee2016posing} have divided the guidelines into body-centered guidelines, object-centered guidelines and space-centered guidelines, which could be seen from Fig.~\ref{fig:guideline}. Body clearance indicates the distance between body and desk, body reachability indicates the limitations of our hands and arms. They also modified the guideline from the general population to a person. The elbow angle for sitting posture is explained between 70 to 100 degrees. 
 
 In this paper, we consider using the person-oriented ergonomic guideline for design evaluation and decision-making. We put the user in the center place, where we first study the needs of users. The user cases are concentrated in order to establish design guidelines. We first find the gap of designing for the general population and designing for minorities. As most users are suffering from various pains (e.g. neck pain, musculoskeletal disorders, etc), which is more or less indicating that there might be improper design problems. Since designer-centered workstation prototype cannot satiate the requirement of all users, we aim to make the design approachable to all users whether they are tall or short, fat or thin, young or old. We are trying to specially take care of minorities and make  design guidelines for the convenience of designers as shown in left portion in Fig.~\ref{fig:design_paradigm}. 
 \subsubsection{ Extending human-centered design }
As shown in the middle portion of Fig.~\ref{fig:design_paradigm}, we try to incorporate the human-centered design paradigm based on the ergonomic design guidelines. The aim is to develop the situated and interactive environment to assist designer who provide various design solutions. We try to initiate a system that first enables user and designer to achieve the customization of workstation, and the system can provide real-time feedback and recommendations to the design and body pose.  Based on the guidelines and needs, we design our system that enables designers to make human-centered office workstation prototypes and provide users with various solutions. For office workstation design, there will be many kinds of types, either for sitting or for standing. Thus, we strive to make the system possible for designers to provide various the digital prototypes and make adjustments while resorting to the preference of users.
\begin{figure*}[t!]
  \centering
  \includegraphics[width=\textwidth, height=8cm]{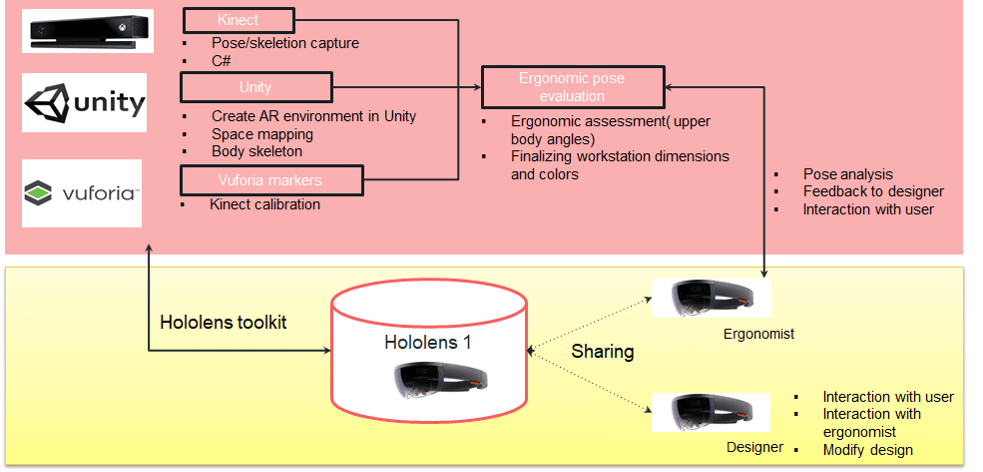}
  \caption{The system diagram of our MR-based interactive design system via augmented multi-person collaboration. Our MR system consists of information sharing module, multi-person interaction module and spatial engagement module. We utilize Kinect, vuforia and Hololens as our hardware components and we develop our MR-based design system in Unity. The information sharing module consist of sharing information among user, designer and ergonomist. The multi-person interaction module contains interaction between user and designer, user and ergonomist, and user and ergonomist. The spatial engagement module contains design analysis and pose analysis. }
  \label{fig:mr_system}
\end{figure*}
\subsubsection{Collaborative evaluation from expert}
As mentioned above, the user and designer can achieve the goal of design and customization, however, whether it is ergonomically reasonable or not should be based on professional ergonomist. As shown in the right portion of Fig.~\ref{fig:design_paradigm}, we initialize new paradigm based on expert inspection, in which the controlled pose study is conducted. 
We focus on the evaluation of ergonomist who will provide more robust analysis for the design and more scientific feedback based on the pose study and estimation. We claim that the collaboration with ergonomist can ensure the design quality based on the prototype from both users and designers. Besides, ergonomist can have deeper analysis on some unreasonable poses and make decisions with designers to ensure the product or prototypes to be comfortable and safe. 
\begin{figure}[h!]
  \centering
  \includegraphics[width=\columnwidth, height=4cm]{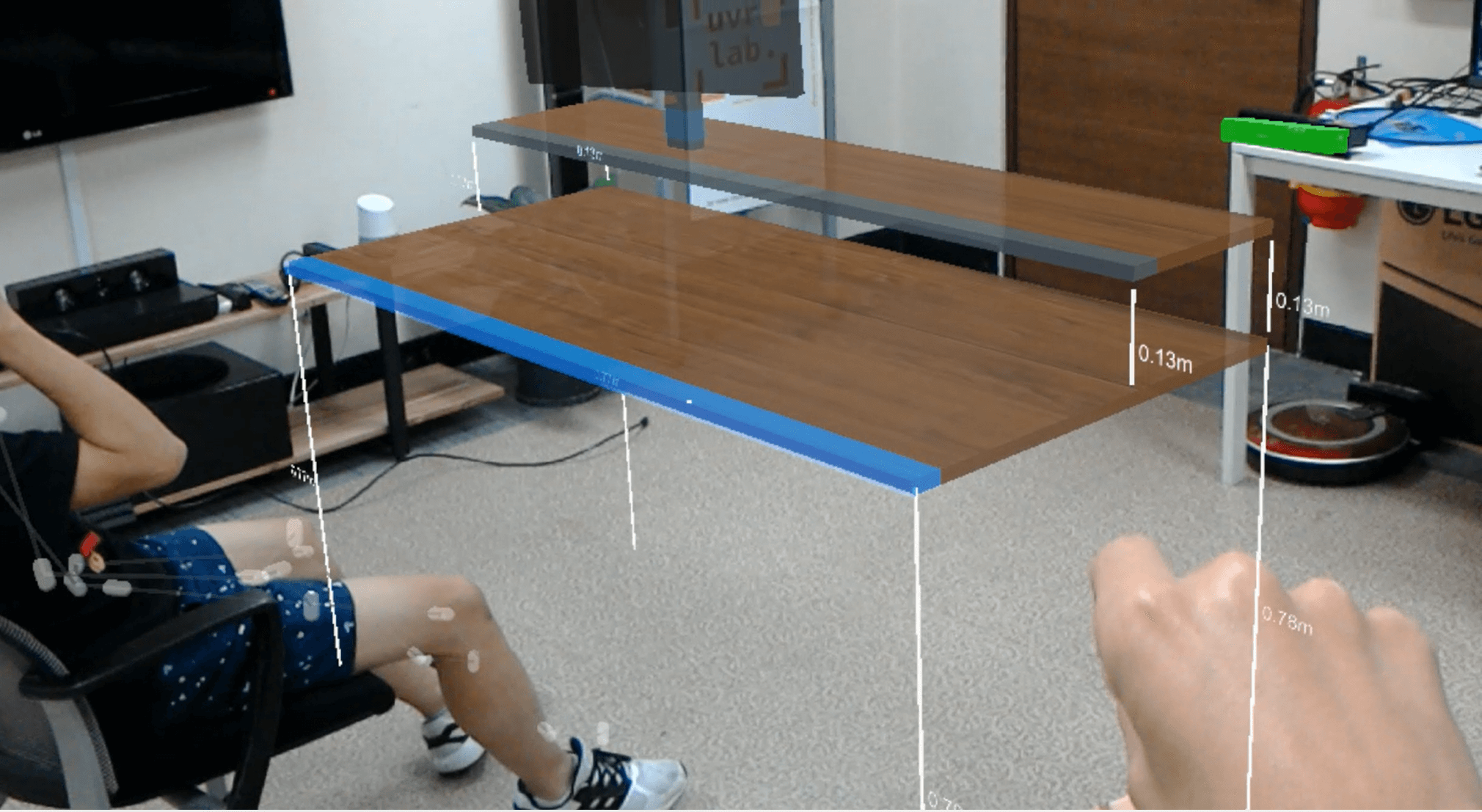}
  \caption{A captured scene of virtual office workstation in physical space from our MR-based design system.}~\label{fig:desk}
  \vspace{-15pt}
\end{figure}
\subsection{MR-based interactive design via augmented collaboration}
Fig.~\ref{fig:mr_system} shows the hardware and software configuration of our MR-based design system. We highlight three crucial interaction modules for our system, namely, information-sharing, multi-user interaction and spatial engagement.
\subsubsection{Information sharing}
\begin{figure}[h!]
  \centering
  \includegraphics[width=\columnwidth, height=4cm]{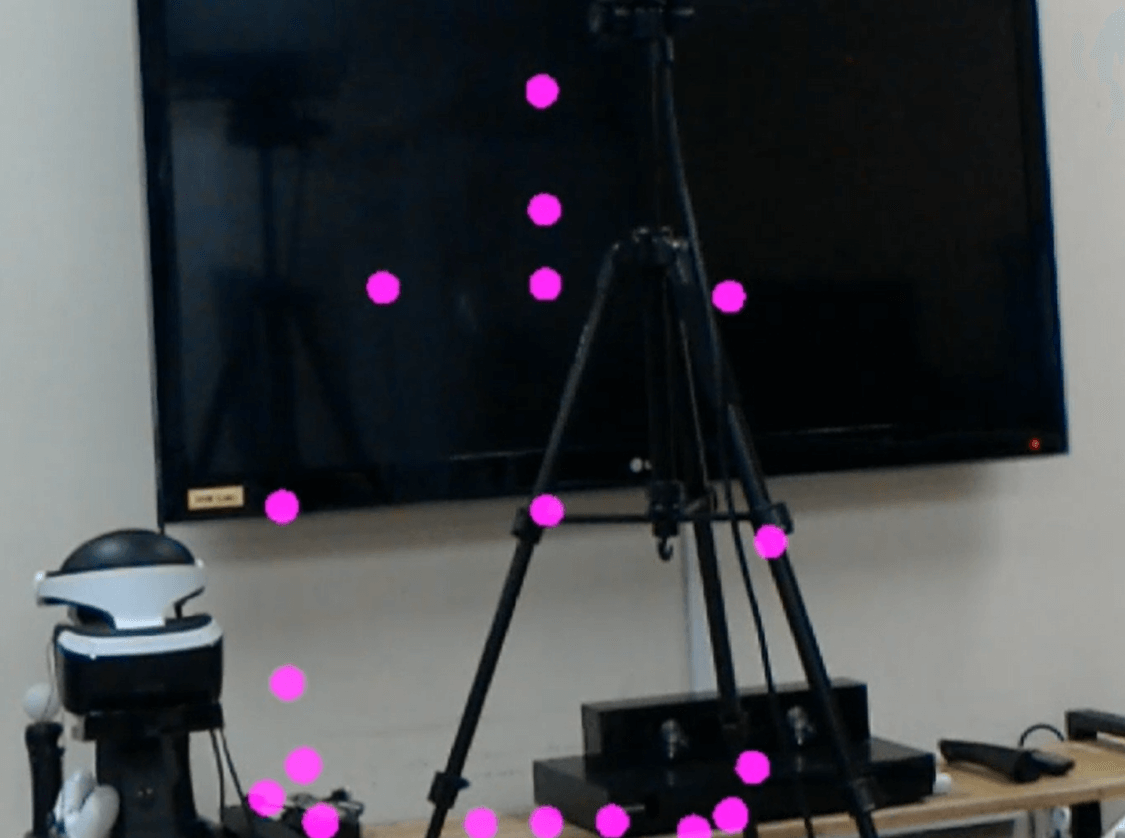}
  \caption{A scene showing the body joints captured from Kinect and projected to Hololens space (MR space).}~\label{fig:pose}
  \vspace{-10pt}
\end{figure}

The ergonomic guidelines and workstation dimensions can be provided in different ways. In this research, we focus on visualizing them in the augmented space. The ergonomic postures and angles from back, neck, and elbow are visualized based on the ergonomic guidelines from \cite{lee2018interactive,pheasant2018bodyspace, vink2004comfort}. We visualize the ergonomic guidelines based on real people in physical space under MR environment and the dimensions (height, width, length, etc.) of virtual workstation as shown Fig.~\ref{fig:desk}. Through the topology mapping of Hololens, the virtual desk is connected with the ground, and also the physical dimensions to the ground is also consistent with real-world dimensions. Besides, upper body angles are highlighted when the angles are beyond the range of recommendations. We aim to make the workstation design a process that includes information gathering, and information collection, especially the user's preference and pose, and eventually information sharing among all participants. Fig.~\ref{fig:pose} shows the body joints from Kinect. We mainly track the upper body joints when user is sitting, and all body joints when standing. We fuse body pose tracking with Hololens projection, such that the pose tracked from real person will be projected to MR space, where the ergonomic information will shared to all people joining the system. We also calculated the RULA \cite{haggag2013real} angles based on the ergonomic guidelines from \cite{lee2018interactive}. In design process, the information related to workstation and RULA angles based ergonomic guidelines will be shared in the MR space.  
\subsubsection{Multi-person interaction}
As the design and evaluation are conducted among user, designer, and ergonomist, and meanwhile, design guidelines are associated with user's body, pose and preference, hence, the system needs to provide communication and interaction between the user, designer and ergonomist. For instance, when the user wants to fix certain desk height after adjustments, the designer needs to get feedback from the user and observe user's body and workstation dimensions to make sure that they are in the allowed ranges or sizes. Meanwhile, the ergonomist and designer need to provide feed-forward and feedback information to each other when the user performs tasks. For the ergonomist and user, they also need to have real-time interaction. For example, when a user says that he/she is content with the workstation design, the egonomist has to check several representative poses and make sure that the ergonomic evaluation is reasonable and reliable. This idea is to make the interaction flows not only between human and computer but also human and human. We accomplish this multi-person interaction via the sharing of Hololens space. As shown in Fig.~\ref{fig:mr_system}, there exists Hololens one ( we call as master Hololens). In reality, the master Hololens will mostly dominated by the user, and there are two slave Hololens that are used by designer and ergonomist. 
\subsubsection{Spatial engagement} 
One crucial aspect for the collaborative design and evaluation is that it needs to support immersive  environment where all people who participate can be fully engaged. One challenge might be the case where user cannot see the virtual scene clearly due to the narrow field of view of Hololens. Therefore, the system must ensure that all information in Fig.~\ref{fig:codesign} should be shown in the augmented space, and all participants could be able to fully engaged in the design process. If all necessary design parameters, related anthropometrics and ergonomic recommendations are not apparently shown, either user or designer will lose concentration, which could be detrimental to the effectiveness of design and evaluation. Upon the fact that Hololens has narrow field of view (FoV), however, we successfully merge all information needed to Hololens space. We initiate some space projection methods based on vuforia markers after Kinect calibration. We then fuse the body skeleton information with the virtual workstation to the MR environment. We show all the information needed for user, designer and ergonomist in the shared space so that they can be engaged to the largest extent. We also conduct usability study, which will be described in next section.
\subsubsection{System configuration}
\begin{figure}[h!]
  \centering
  \includegraphics[width=\columnwidth, height=8cm]{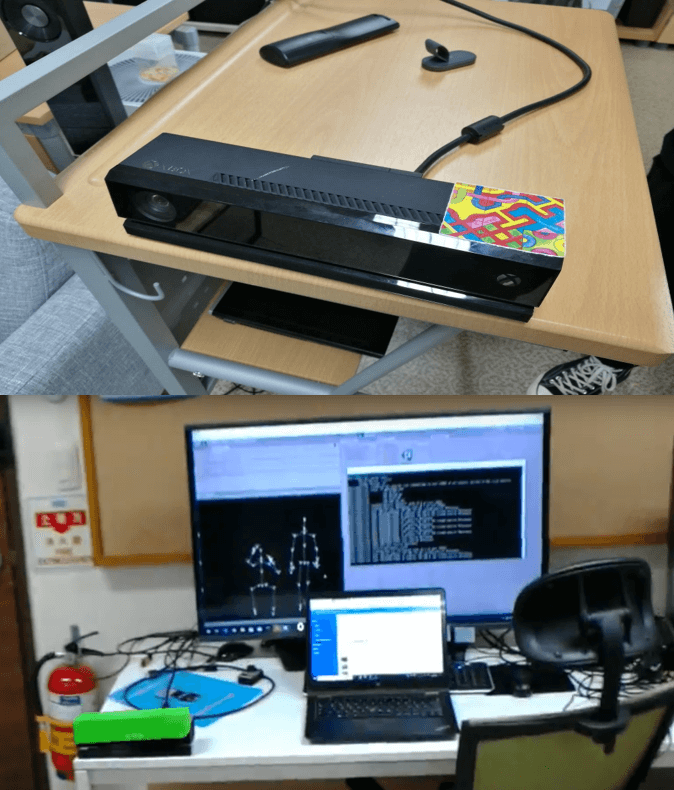}
  \caption{A scene showing vuforia marker for localization and calibration. The upper figure shows the vuforia marker we designed for localization of Kinect. Bottom one shows that Kinect is projected to Hololens space as a green box.  } 
  \label{fig:settings}
  \vspace{-10pt}
\end{figure}

Fig.~\ref{fig:mr_system} depicts the detailed principle of our system. For hardware, it contains Kinect, vuforia makers, Hololens and PC (Windows 10, 8G memory). For software, it consists of Unity together with $\texttt{C\#}$ API. The configuration is built in Unity 3D as a platform for creating the basic MR environment. We create two types of office workstations by a designer, which are somehow simplified but similar to the real ones. A virtual computer with keyboard and mouse was placed in the workstation so that it provides the user with the chance to be fully immersed when performing tasks as shown in \ref{fig:desk}.  The system consists of three key functions. \emph{First of all}, in order to conduct an ergonomic evaluation, we use Microsoft Kinect as the skeleton and pose tracking device. The main aim is to track body joints and conduct ergonomic evaluation using RULA\cite{pontonnier2014designing} based on the guidelines from \cite{lee2016posing} for upper body estimation. The measured angles include viewpoint angles from eyes, arm angles, trunk flexion angle, and lower arm flexion angle. These angles and poses in ergonomic measurements are further projected to MR space generated in Hololens. \emph{Second}, calibration and localization using vuforia marker as shown in the top portion of Fig.~ \ref{fig:settings}. The calibration is done through a virtual box created in MR space and is based on the affine transformation matrix calculated from quaternion as shown in the bottom portion of Fig.~\ref{fig:settings}. Besides, the vuforia marker is also utilized as an approach to localize Kinect in MR space. Since the pose tracked by Kinect from a user is aligned with coordinate of Hololens, through localization, it is possible to share all topology information. \emph{Lastly}, the distance of office workstation layer to ground and workstation dimensions are shared in the MR space, where user, designer and ergonomist can view and refer. 
The ergonomic body pose evaluation was assessed in Hololens space, where the projected body joints and measured angles from Kinect are shown in the user's body. Meanwhile, the virtual desk is displayed in front of the user. The sharing of Hololens scenes is conveyed through a server, which connects other Hololens through WiFi. The collaborative design system aims to personalize workstation design with ergonomic guidelines through the incorporation with anthropometrics, pose and design dimensions through sharing in MR space. The system not only focuses on the interaction between human and computer but also the interaction between human and human in the augmented space. 
\begin{figure*}[t!]
  \centering
  \includegraphics[width=0.95\textwidth, height=4.5cm]{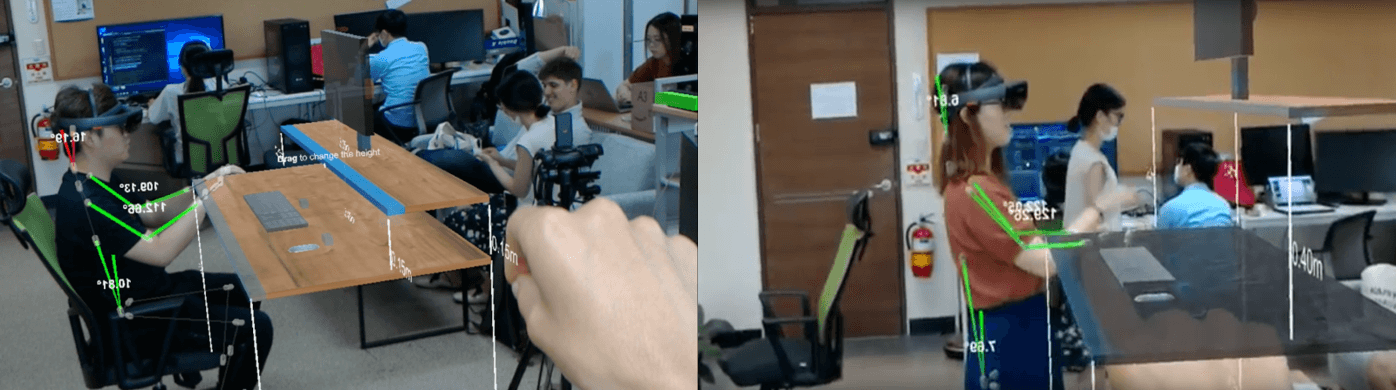}
  \caption{Experimental results of our proposed MR-based design system via augmented collaboration. Left shows the workstation design and evaluation when user is sitting, and right shows the scene when user is standing. The design information (workstation dimensions and color, etc) and ergonomic information (RULA angles from ergonomic guidelines) are shared with all participants. }~\label{fig:results}
  \vspace{-10pt}
\end{figure*}

\section{User Study and Evaluation}
In order to evaluate our system, we conducted usability study and tried to get some feedback from participants about how they thought about our MR-based design system. We are interested in how well they feel about the non-specialists could collaborate with the designers and to what extent they could feel the augmented workstation when performing tasks. During the study, we collected preliminary qualitative feedback to gain insight into the experience of collaborating with the designer and ergonomist.
\subsection{Participants}
We recruited one designer who was a student majoring industrial design (Age:26, M) and had experience of product design using VR. We also recruited one researcher majoring in human factors (Age:28, M) and he had experience of using MR devices. We also recruited 5 participants as users between the ages of 25 to 32 (M: 26.4, SD=1.95) to do a preliminary evaluation of our system. The users all had no experience in industrial design or had no knowledge about ergonomics. Most of them never studied the principle of human-computer interaction and VR/AR. Most participants worked mostly at school library or labs. All participants had their own preference and needs for their office workstations.
\subsection{Procedure of usability study}
The usability study consists of three stages: familiarization of design and interaction, collaborative design, and feedback survey. The goal of the first stage was to let the participants know what was the hardcore of design and the basics of human-computer interaction, especially interaction in MR. 
We started by explaining the concept of CoAug-MR in conjunction with the interface, interactions and design strategies. The participants were first asked to listen to our introduction to the tasks and system principles. We introduced why we did this, what they can see from Hololens and what they will be doing in the augmented environment.
In the second stage, the participants were asked to join the design with experts.We first provided a chance for them to practice, then the users started designing with the designer and ergonomist until all them are satisfied with their expectation. 
The design was immediately followed by an questionnaire survey in the third stage regrading their design and engagement in the design and evaluation. For analysis, we checked the feedback from the users from each question and also the open suggestions in the last question.

\subsection{Results and discussion }
All of the users participated in the design of office workstation that met their needs and preferences via augmented multi-person collaboration.   
As shown in the left of Fig.\ref{fig:results}, one subject was designing the workstation with the designer in augmented space locally, while ergonomist was checking the body poses and provideed ergonomic analysis based on design. The right side of in Fig.~\ref{fig:results} shows the scene where the user was designing the workstation with the designer in standing pose locally, and ergonomist was also conducting ergonomic evaluation. The ergonomic pose angles and office workstation information were all shared in the MR space. The user, designer and ergonomist will interact and instantly achieve the design task via augmented collaboration.
Right after the users' participation of design with designer and ergonomist, they were asked to immediately finish a questionnaire survey. 

\begin{figure}[h!]
  \centering
  \includegraphics[width=0.95\columnwidth, height=3.5cm]{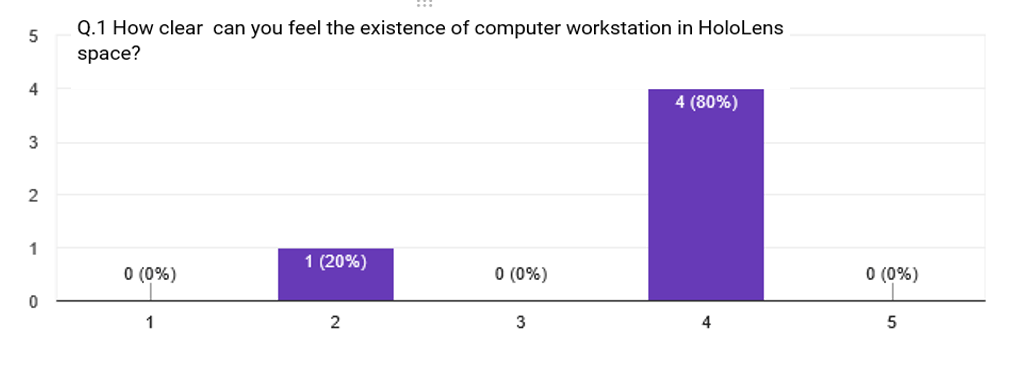}
  \caption{The usability study results from question.1. Around 20\% of participants respond that they were okay to see the virtual office workstation in MR space even though they felt the FoV of Hololens was narrow. And 80\% participants were satisfied with the clearance of office workstation.}~\label{fig:Q1}
  \vspace{-10pt}
\end{figure}

The first question we asked was `how clearly can you feel the existence of computer workstation in Hololens space?'. As shown in Fig.~\ref{fig:Q1}, a few  participants replied that the field of view of Hololens was narrow, but it did not affect the clear existence of office workstation in the Hololens. And 80\% of them  were very satisfied with the 3D office workstation model in Hololens. 

\begin{figure}[h!]
  \centering
  \includegraphics[width=0.95\columnwidth, height=3.5cm]{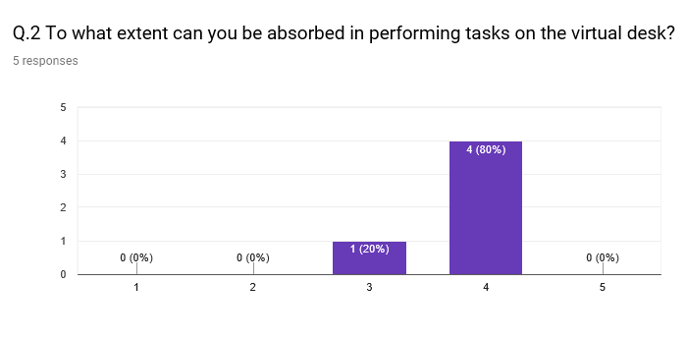}
  \caption{Usability study results from question 2. Most participants respond that they are fully absorbed in the design process.}~\label{fig:Q2}
  \vspace{-10pt}
\end{figure}

In the following question `to what extent can you be absorbed in performing the task on the virtual desk?', we asked them about their concentration when performing the design tasks. As shown in Fig.~\ref{fig:Q2}, 20\% of participants replied that they were almost immersed when performing design tasks. Most users (80\%) were fully concentrated on the design and prototyping of the office workstation, and had a nice interaction with both designer and ergonomist via the augmented collaboration.   

\begin{figure}[h!]
  \centering
  \includegraphics[width=0.95\columnwidth, height=3.5cm]{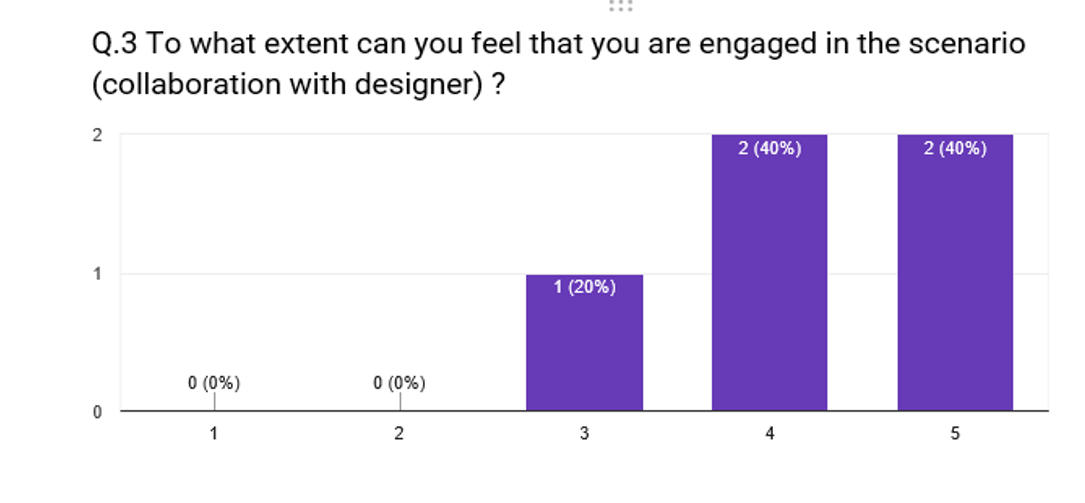}
  \caption{Usability study results from question 3. Most participants (around 80\%) were highly confident that they were able to be engaged in the collaboration with designer and ergonomist. }~\label{fig:Q3}
  \vspace{-10pt}
\end{figure}

In question 3, we asked `to what extent they felt engaged during the collaboration with the designer in certain scenarios?'. 80\% of participants replied that they were highly content with the engagement with the designer as shown in Fig.~\ref{fig:Q3}. They mentioned that the design was interesting since they can have a real-time design under the assistance of designer and ergonomist, who provided professional knowledge for them regarding office workstation style and safe body pose. This actually reflects that our MR system is well designed for engagement of users, designer and ergonomist. 

\begin{figure}[h!]
  \centering
  \includegraphics[width=0.95\columnwidth, height=3.5cm]{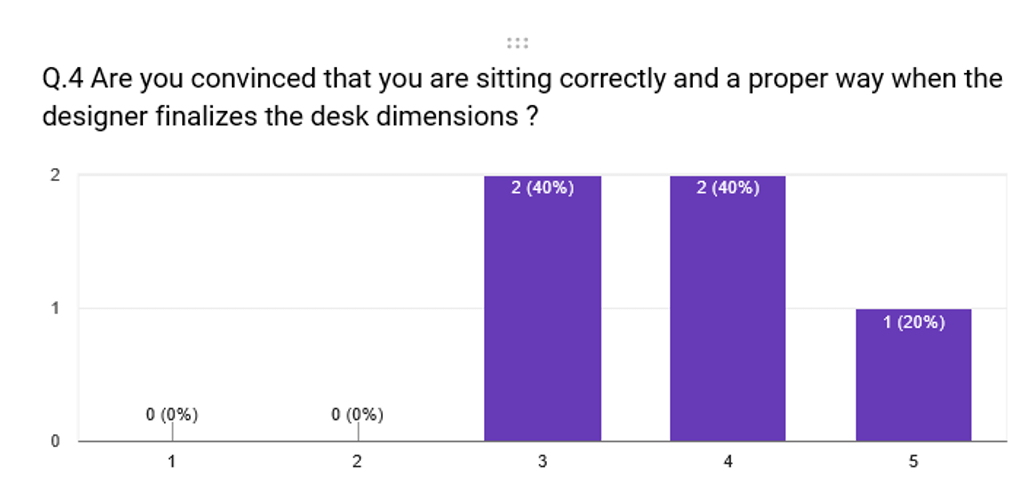}
  \caption{Usability study results from question 4. More than 60\% participants replied that they were highly convinced that they were sitting in a proper way, which was consistent with the feedback from ergonomist.}~\label{fig:Q4}
 \vspace{-10pt}
\end{figure}

We then evaluated how much they were convinced about their poses when they said they were sitting in a proper way or standing correctly. As shown in Fig.~\ref{fig:Q4}, more than 60\% of the participants said that they were highly sure that their poses were good for them when they finalized their expected workstation dimensions done by the designer. The remained replied they are almost sure that they were sitting in the proper position in a correct pose. The results show that our system can accurately estimate their body poses, which is more efficient (no need for DHMs and  mannequins) and reliable for dangerous pose prevention. 

\begin{figure}[h!]
  \centering
  \includegraphics[width=0.95\columnwidth, height=3.5cm]{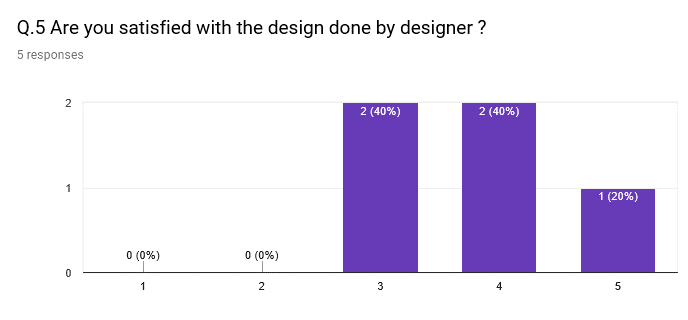}
  \caption{Usability study results from question 5. Most people are satisfied with the design work done with designer.}~\label{fig:Q5}
  \vspace{-10pt}
\end{figure}

In question 5, we asked them about their satisfaction with the design of office workstation made by the designer. Fig.~\ref{fig:Q5} shows that more than 60\% of participants gave a reasonable feedback about the design done by the designer. The other two participants said that they were somehow sure about the design quality due to the small field of view of Hololens. \\

Lastly, we provided that chance for them to give any suggestions to the design of our system. Most people said the design is fabulous, however, the field of view of Hololens was too narrow, which might weaken the quality of design and evaluation experience in the experiment. They all wanted to get a wider view for better immersion in the augmented space. In the near future, we will find an alternative to improve the effectiveness of design engagement.

\section{ Conclusion and Future Work}
In this paper, we have presented CoAug-MR, a MR-based interactive design method via augmented multi-person collaboration. We emphasized the importance of collaboration, and initiate an augmented design paradigm, which enables user, designer and even ergonomist to gather up to achieve reliable and real-time design and prototyping based on user's pose either locally or remotely. The evaluation of the system showed that users can successfully conduct collaborative design with designer based on ergonomic evaluations with experts, which demonstrated the effectiveness of our proposed method. It is worthy of mentioning that our system is also capable of designing any other products based on the usage and application, which will be our future work. We will also focus more on remote collaboration via teleportation technology, which provides the possibility to transport a person or a product instantly from one place to the other. In this way, our proposed CoAug-MR can handle more complex product design and remote collaboration problem. The other goal is to find better device for better user experience since Hololens has some drawbacks in terms of spatial engagement.     

\balance{}

\bibliographystyle{SIGCHI-Reference-Format}
\bibliography{sample}


\begin{thebibliography}{00}


\ifx \showCODEN    \undefined \def \showCODEN     #1{\unskip}     \fi
\ifx \showDOI      \undefined \def \showDOI       #1{{\tt DOI:}\penalty0{#1}\ }
  \fi
\ifx \showISBNx    \undefined \def \showISBNx     #1{\unskip}     \fi
\ifx \showISBNxiii \undefined \def \showISBNxiii  #1{\unskip}     \fi
\ifx \showISSN     \undefined \def \showISSN      #1{\unskip}     \fi
\ifx \showLCCN     \undefined \def \showLCCN      #1{\unskip}     \fi
\ifx \shownote     \undefined \def \shownote      #1{#1}          \fi
\ifx \showarticletitle \undefined \def \showarticletitle #1{#1}   \fi
\ifx \showURL      \undefined \def \showURL       #1{#1}          \fi

\bibitem{anderson2013youmove}
{Fraser Anderson}, {Tovi Grossman}, {Justin Matejka}, {and} {George
  Fitzmaurice}. 2013.
\newblock \showarticletitle{YouMove: enhancing movement training with an
  augmented reality mirror}. In {\em Proceedings of the 26th annual ACM
  symposium on User interface software and technology}. ACM, 311--320.
\newblock


\bibitem{berg2017industry}
{Leif~P Berg} {and} {Judy~M Vance}. 2017.
\newblock \showarticletitle{Industry use of virtual reality in product design
  and manufacturing: a survey}.
\newblock {\em Virtual reality\/} {21}, 1 (2017), 1--17.
\newblock


\bibitem{follmer2010copycad}
{Sean Follmer}, {David Carr}, {Emily Lovell}, {and} {Hiroshi Ishii}. 2010.
\newblock \showarticletitle{CopyCAD: remixing physical objects with copy and
  paste from the real world}. In {\em Adjunct proceedings of the 23nd annual
  ACM symposium on User interface software and technology}. ACM, 381--382.
\newblock


\bibitem{grecco2007virtual}
{Claudio Henrique dos~Santos Grecco}, {Isaac Jos{\'e} Ant{\^o}nio Luquetti~dos
  Santos}, {Ant{\^o}nio Carlos~Abreu Mol}, {Paulo Victor Rodrigues~de
  Carvalho}, {Antonio~Carlos Silva}, {Francisco Jos{\'e}~Oliveira Ferreira},
  {Marco Dutra}, {and} {others}. 2007.
\newblock \showarticletitle{Virtual reality technology as a tool for human
  factors requirements evaluation in design of the nuclear reactors control
  desks}.
\newblock  (2007).
\newblock


\bibitem{haggag2013real}
{Hussein Haggag}, {Mohammed Hossny}, {Saeid Nahavandi}, {and} {D Creighton}.
  2013.
\newblock \showarticletitle{Real time ergonomic assessment for assembly
  operations using kinect}. In {\em 2013 UKSim 15th International Conference on
  Computer Modelling and Simulation}. IEEE, 495--500.
\newblock


\bibitem{ikonomov2003using}
{Pavel~G Ikonomov} {and} {Emiliya~D Milkova}. 2003.
\newblock \showarticletitle{Using virtual reality simulation through product
  lifecycle}. In {\em ASME 2003 International Mechanical Engineering Congress
  and Exposition}. American Society of Mechanical Engineers Digital Collection,
  761--767.
\newblock


\bibitem{lee2016posing}
{Bokyung Lee}, {Minjoo Cho}, {Joonhee Min}, {and} {Daniel Saakes}. 2016.
\newblock \showarticletitle{Posing and acting as input for personalizing
  furniture}. In {\em Proceedings of the 9th Nordic Conference on
  Human-Computer Interaction}. ACM, 44.
\newblock


\bibitem{lee2018interactive}
{Bokyung Lee}, {Joongi Shin}, {Hyoshin Bae}, {and} {Daniel Saakes}. 2018.
\newblock \showarticletitle{Interactive and Situated Guidelines to Help Users
  Design a Personal Desk That Fits Their Bodies}. In {\em Proceedings of the
  2018 Designing Interactive Systems Conference}. ACM, 637--650.
\newblock


\bibitem{mattmann2007recognizing}
{Corinne Mattmann}, {Oliver Amft}, {Holger Harms}, {Gerhard Troster}, {and}
  {Frank Clemens}. 2007.
\newblock \showarticletitle{Recognizing upper body postures using textile
  strain sensors}. In {\em 2007 11th IEEE international symposium on wearable
  computers}. IEEE, 29--36.
\newblock


\bibitem{mochimaru2017digital}
{Masaaki Mochimaru}. 2017.
\newblock \showarticletitle{Digital human models for human-centered design}.
\newblock {\em Journal of Robotics and Mechatronics\/} {29}, 5 (2017),
  783--789.
\newblock


\bibitem{morisawa2003development}
{Toshihiro Morisawa}, {Masashi Sawa}, {Kenji Tamaki}, {Toshiharu Miwa},
  {Kousaku Tachikawa}, {Hisahiko Abe}, {Toshihiro Nakajima}, {and} {Yoshio
  Iwata}. 2003.
\newblock \showarticletitle{Development of Run-to-Run Control System for Oxide
  Film CMP for High-Product Mix Fab, 4th European AEC}. In {\em APC
  Conference}.
\newblock


\bibitem{noh2015hmd}
{Seung-Tak Noh}, {Hui-Shyong Yeo}, {and} {Woontack Woo}. 2015.
\newblock \showarticletitle{An HMD-based mixed reality system for
  avatar-mediated remote collaboration with bare-hand interaction}. In {\em
  Proceedings of the 25th International Conference on Artificial Reality and
  Telexistence and 20th Eurographics Symposium on Virtual Environments}.
  Eurographics Association, 61--68.
\newblock


\bibitem{ottosson2002virtual}
{Stig Ottosson}. 2002.
\newblock \showarticletitle{Virtual reality in the product development
  process}.
\newblock {\em Journal of Engineering Design\/} {13}, 2 (2002), 159--172.
\newblock


\bibitem{park2008design}
{Hyungjun Park}, {Jeong-Soo Son}, {and} {Kwan-Heng Lee}. 2008.
\newblock \showarticletitle{Design evaluation of digital consumer products
  using virtual reality-based functional behaviour simulation}.
\newblock {\em Journal of Engineering Design\/} {19}, 4 (2008), 359--375.
\newblock


\bibitem{pastura2013joint}
{FCH Pastura}, {GL de Almeida}, {and} {G CUNHA}.
\newblock \showarticletitle{Joint angles calculation through augmented
  reality}. In {\em Proceedings of The 2nd International Digital Human Modeling
  Symposium 2013}.
\newblock


\bibitem{peng2007virtual}
{Qingjin Peng}. 2007.
\newblock \showarticletitle{Virtual Reality Technology in product design and
  manufacturing}.
\newblock {\em Proceedings of the Canadian Engineering Education Association
  (CEEA)\/} (2007).
\newblock


\bibitem{pheasant2018bodyspace}
{Stephen Pheasant} {and} {Christine~M Haslegrave}. 2018.
\newblock {\em Bodyspace: Anthropometry, ergonomics and the design of work}.
\newblock CRC Press.
\newblock


\bibitem{pontonnier2014designing}
{Charles Pontonnier}, {Georges Dumont}, {Asfhin Samani}, {Pascal Madeleine},
  {and} {Marwan Badawi}. 2014.
\newblock \showarticletitle{Designing and evaluating a workstation in real and
  virtual environment: toward virtual reality based ergonomic design sessions}.
\newblock {\em Journal on Multimodal User Interfaces\/} {8}, 2 (2014),
  199--208.
\newblock


\bibitem{saul2011sketchchair}
{Greg Saul}, {Manfred Lau}, {Jun Mitani}, {and} {Takeo Igarashi}. 2011.
\newblock \showarticletitle{SketchChair: an all-in-one chair design system for
  end users}. In {\em Proceedings of the fifth international conference on
  Tangible, embedded, and embodied interaction}. ACM, 73--80.
\newblock


\bibitem{schumann1998applying}
{Hagen Schumann}, {Silviu Burtescu}, {and} {Frank Siering}. 1998.
\newblock \showarticletitle{Applying augmented reality techniques in the field
  of interactive collaborative design}. In {\em European Workshop on 3D
  Structure from Multiple Images of Large-Scale Environments}. Springer,
  290--303.
\newblock


\bibitem{taylor2013posture}
{Brett Taylor}, {Max Birk}, {Regan~L Mandryk}, {and} {Zenja Ivkovic}. 2013.
\newblock \showarticletitle{Posture training with real-time visual feedback}.
\newblock  (2013).
\newblock


\bibitem{vink2004comfort}
{Peter Vink}. 2004.
\newblock {\em Comfort and design: principles and good practice}.
\newblock CRC press.
\newblock


\bibitem{wang2016zishi}
{Qi Wang}, {Marina Toeters}, {Wei Chen}, {Annick Timmermans}, {and} {Panos
  Markopoulos}. 2016.
\newblock \showarticletitle{Zishi: a smart garment for posture monitoring}. In
  {\em Proceedings of the 2016 CHI Conference Extended Abstracts on Human
  Factors in Computing Systems}. ACM, 3792--3795.
\newblock


\bibitem{wegner2007digital}
{Diana Wegner}, {Jim Chiang}, {Brent Kemmer}, {Dan L{\"a}mkull}, {and} {Roland
  Roll}. 2007.
\newblock {\em Digital human modeling requirements and standardization}.
\newblock {T}echnical {R}eport. SAE Technical Paper.
\newblock


\bibitem{weichel2014mixfab}
{Christian Weichel}, {Manfred Lau}, {David Kim}, {Nicolas Villar}, {and}
  {Hans~W Gellersen}. 2014.
\newblock \showarticletitle{MixFab: a mixed-reality environment for personal
  fabrication}. In {\em Proceedings of the SIGCHI Conference on Human Factors
  in Computing Systems}. ACM, 3855--3864.
\newblock


\bibitem{wu2013usability}
{Hsin-Chieh Wu}, {Min-Chi Chiu}, {Cheng-Lung Lee}, {and} {Ming-Yao Bai}. 2013.
\newblock \showarticletitle{Usability evaluation of the universal computer
  workstation under supine, sitting and standing postures}. In {\em
  International Conference on Human Interface and the Management of
  Information}. Springer, 151--156.
\newblock


\bibitem{ye2007applications}
{Jilin Ye}, {Saurin Badiyani}, {Vinesh Raja}, {and} {Thomas Schlegel}. 2007.
\newblock \showarticletitle{Applications of virtual reality in product design
  evaluation}. In {\em International Conference on Human-Computer Interaction}.
  Springer, 1190--1199.
\newblock


\bibitem{yee2009augmented}
{Brandon Yee}, {Yuan Ning}, {and} {Hod Lipson}. 2009.
\newblock \showarticletitle{Augmented reality in-situ 3D sketching of physical
  objects}. In {\em Intelligent UI workshop on sketch recognition}, Vol.~1.
  Citeseer.
\newblock


\end{thebibliography}

\end{document}